\documentstyle{article}

\catcode`\@=11

\def\slash{\global\setbox0=\hbox{\raise.075em
     \hbox{\m@th\kern-.075em $\mathchar"0236$}}%
     \wd0=0pt \ht0=0pt \dp0=0pt \box0}

\catcode`\@=12
\newcommand{\be}{\begin{equation}}
\newcommand{\ee}{\end{equation}}
\begin{document}
\title{Kaplan-Narayanan-Neuberger lattice fermions pass a
perturbative test\thanks{Princeton University Preprint PUPT-1510;
hep-th/9411137}}
\author{ S. Aoki\thanks{saoki@het.ph.tsukuba.ac.jp} \\{\it
Institute of Physics, University of Tsukuba,}
\\ {\it Tsukuba Ibaraki-305, Japan}
\\R.B. Levien\thanks{levien@pupgg.princeton.edu} \\{\it
Department of Physics, Joseph Henry Laboratories,}
\\ {\it Princeton University , Princeton, N.J. 08544}
}
\date{17 November 1994}
\maketitle
\begin{abstract}
We test perturbatively a recent scheme for implementing chiral
fermions on the lattice, proposed by Kaplan and modified by
Narayanan and Neuberger, using as our testing ground
the chiral Schwinger model. The scheme is found to reproduce the
desired form of the effective action, whose real part is gauge
invariant and whose imaginary part gives the correct anomaly in
the  continuum limit, once technical problems relating to the
necessary infinite extent of the extra dimension are properly
addressed. The indications from this study are that the
Kaplan--Narayanan--Neuberger (KNN) scheme has a good chance at
being a correct lattice regularization of chiral gauge theories.
\end{abstract}
\section{Introduction}
There has been much progress recently in an old problem in
the understanding of gauge theories, namely the regularization of chiral gauge
theories. The goal is a gauge invariant regularization: while in theory there
is nothing wrong with regulators breaking gauge invariance, we would like a
gauge invariant regularization for at least two distinct reasons. In
perturbation theory a gauge invariant regularization makes the proof of
renormalizability much simpler\cite{Fr:Sl:93,Fa:Sl:89}. For non-perturbative
calculation much of the success of lattice field theory has followed
directly from its manifest gauge invariance, so we are reluctant to throw this
away. Lattice regularization of a chiral theory however must be clever enough
to evade no-go theorems\cite{Ka:Sm:81,Ni:Ni:81} which state that it is
impossible to have simultaneously (1) locality, (2) chiral invariance, (3) the
correct number of fermion species.

A good overview of the problem has been provided by Narayanan and
Neuberger\cite{Na:Ne:93}, who point out that two of the most promising recent
schemes, one perturbative\cite{Fr:Sl:93} and another on the lattice\cite{Ka:92}
, both make use of a common trick.
A theory which looks vector like is constructed by coupling right-handed
particles to a mass matrix $M$ and left-handed particles to $M^\dagger$, as in
the following (Euclidean) Lagrangian:
\be
{\cal L} = - \bar{\psi} \slash D \psi + \bar{\psi} (M P^R + M P^L) \psi,
\label{Lchiral}
\ee
where $P^R = \frac{1}{2}(1 + \gamma_5)$, $P^L = \frac{1}{2}(1-\gamma_5)$, and
$\slash D$ is the covariant derivative.  For
the theory to describe a right-handed fermion we need $M$ to have a zero mode
while $M^\dagger$ has none: thus the mass matrix $M$ needs to have infinite
dimension. The infinite number of extra fields are realised as Pauli-Villars
regulators in reference\cite{Fr:Sl:93}; in reference \cite{Ka:92} the mass
matrix is realised by a domain wall in a higher dimension, labelled by
coordinate $s$.  There is a right
handed zero mode bound to the domain wall.

It is clear from the above that in the domain wall scheme the extent of the
extra dimension should be kept infinite to avoid creating fermions in pairs of
opposite chirality. Indeed if the extra dimension is made finite we have a
anti-domain wall, to which a left-handed zero mode is bound. One
can then try
to restrict the gauge fields to a ``waveguide'' around the
domain wall, with scalar fields inserted at the boundaries of the
waveguide to
restore gauge invariance: this approach is still under
investigation but all indications are that the theory remains
vector-like\cite{Go:et:al:93-94}. So instead we
follow the rather different approach of Narayanan and
Neuberger\cite{Na:Ne:93,Na:Ne:94}, keeping the $s$-extent
infinite. In the language of the previous paragraph this gives an
infinite dimensional mass matrix (whose explicit form is given in
the next section): in ``domain wall'' language we eliminated the
possibility of zero modes bound to the anti-domain wall, because
now there simply {\em is} no anti-domain wall.
There is no need to introduce new gauge fields, so the gauge
fields have the dimension of the lower dimensional
target space
(they have no $s$-dependence, and are simply copied
to each $s$ slice).

That we have at least the possibility of evading the no-go theorems is
evident: from the point of view of references\cite{Ni:Ni:81,Ka:Sm:81}, if
we imagined integrating out all the extra fermion species except the
righthanded fermion at $s=0$, the action we get is no longer
local\cite{Na:Ne:93}. Other non-local formulations have been
tried but these all either break Lorentz invariance or
dynamically generate ghost contributions which wreck the theory
 (for an excellent treatment of these problems see \cite{Pe:88}).
In this approach the ghosts are cancelled by pure gauge terms
which also come from integrating out the fermions\cite{Na:Ne:93}.

The purpose of this paper is to carry out a test of the
Kaplan-Narayanan-Neuberger (KNN) scheme in
perturbation theory, using as our testing ground an exactly solvable model, the
Schwinger model ($1+1$ dimensional QED). To be more precise, we examine a
chiral Schwinger model that is exactly the usual Schwinger model ``cut in
half'': that is, we have a single right-handed fermion, minimally coupled to
the gauge field. In the continuum limit, we expect the model to be anomalous:
the effective gauge field action induced by
integrating out the fermions should be gauge-variant because of
this anomaly.
This is of no concern for our purposes: to make an anomaly-free
model we have many
options, one fairly routine, but the others directly
relevant to the central problem in
the field, that of regularizing theories like the Standard Model.
The routine
option is to
introduce a left-handed particle to make the theory vector-like: the
left-handed anomaly then has the opposite sign (we check this explicitly), so
that the theory is overall anomaly-free. We note that even the
construction of this theory would be problematic if (as with
other lattice regularizations of chiral theories) our effective action had a
gauge-variant real part. A more interesting choice is to
introduce two right-handed particles, and a left-handed
particle, with couplings to the gauge field in a ratio 3:4:5
(hence ``3-4-5 model"), so that again the anomalies (which are
proportional to the charge squared) cancel. In two dimensions this
is the simplest ``toy" Standard Model, with non-trivial anomaly
cancellation, but clearly we could
construct many other such models.

The point is that because in the KNN formalism each new type of particle
(by ``type'' we mean flavour or chirality) is completely
decoupled from the others, unless we introduce an explicit mass term coupling
them. So in the massless limit there are no ``type-changing'' vertices in
perturbation theory, and we can calculate the anomalies for each type of
particle separately. The essential problem is then to calculate the effective
action and the anomaly for a single flavour and
chirality of fermion, minimally coupled to the gauge field, knowing that for a
model containing only massless fermions we can simply add the effective actions
for each type of fermion together, so that in particular we can make the
anomaly-free theories described above.

In this paper we calculate the one-loop
effective action for two external gauge
fields induced by integrating out the fermions, i.e. the one-loop vacuum
polarization graph. We also show that in the continuum limit fermion
loops with more than two external gauge fields attached do not
survive. This allows us to sum bubble graphs and get results
accurate to all orders of perturbation theory, just as in the
continuum theory. We can thus give
exact expressions for the anomaly, the mass gap and the chiral order
parameter in the continuum limit. We then
compare our results with those obtained by
other regularization methods.

As was pointed out in reference
\cite{Na:Ne:93}, the fermion propagator has the right structure near zero
momentum by construction, so there are only two ways in which the scheme can
fail this perturbative test: firstly
because of the new infinity in
the theory, the infinite $s$-extent necessary to create a genuinely chiral
theory ( a new and not so well understood problem) and secondly
because of the peculiarities of momentum integration on the lattice (an
old
and well understood problem). Narayanan and
Neuberger\cite{Na:Ne:93,Na:Ne:94} have given a prescription for handling the
first problem, the
new infinity: they point out that it is a bulk effect in $s$-space
(this is obvious in their
overlap formalism and in perturbation theory it will become
clear from the fact that
only the translationally invariant part of the propagator diverges) so that it
is naturally cured by subtracting diagrams for which the domain wall mass term
has been replaced by a constant mass.
It now needs to be checked that this scheme
can be implemented without introducing new
singularities which might alter the continuum limit of the
theory. To deal with the
second problem, we need to be careful about taking the continuum
limit $a \rightarrow 0$ of Feynmann integrals (where $a$ is the
lattice spacing). This is because propagators depend on the loop momenta
$q_\mu$ through $q_\mu
a$, which can be of order $1$ since the momenta range from
$-\pi/a$ to $\pi/a$. So a simple expansion of the integrand in
powers of $a$ is not valid. We follow reference\cite{Ka:Sm:81}
and divide the integration region into an ``inside" region near the
propagator pole at zero momentum, and an ``outside" region which
is the rest of the Brioullin zone. It turns out that for fermion
loops with more than two gauge fields attached, only the ``inside"
region (where we can replace the propagators by their continuum
limits) contributes in the continuum limit.
The inside region in turn vanishes because of
Ward identities constraining the continuum
propagators\cite{Bo:Ko:87}. For the remaining graph with two gauge
fields attached (the vacuum polarization), both inside and
outside regions contribute.

In section \ref{subsec} we carry out the $s$-subtractions for the
effective action, and show that they render the initially
$s$-divergent action $s$-finite. In order to make the ill-defined infinite $s$
summation well-defined  we first
restrict the gauge interaction to a finite range $-L \leq s \leq
L$,
while the fermion fields propagate in infinite $s$--space.
We then take the limit $L \rightarrow \infty$ after subtractions.
We emphasise
strongly however that the limit $L \rightarrow \infty$ has to
be taken, and that the
fermion lives in an infinite $s$--space, as we will see in section 5. A
subtle point that needs to be addressed relates to an
ambiguity in the imaginary part of the effective action,
seen in the
overlap formalism\cite{Na:Ne:94}. Because the imaginary part of the
Euclidean action corresponds to the parity-violating part of the action in
Minkowski space, it is the most interesting part, giving rise to the
anomaly for example. In reference \cite{Na:Ne:94} the ambiguity arose
because of an
ambiguous phase in the boundary states, when the effective action was rewritten
as an overlap using transfer matrices in the $s$-direction.
Does such an ambiguity occur in
perturbation theory ?  In
sections \ref{subsec} and \ref{homreal} we show that in perturbation theory
 the imaginary part of
the effective action is finite {\em before} subtractions, and unaffected by the
subtractions (i.e. the subtracted terms are purely real). Thus the imaginary
part of the effective action is {\em unambiguous} in our perturbation scheme.
Obviously the perturbation scheme has picked a phase for the boundary states:
in section \ref{pertphase} it becomes clear that in
our perturbation scheme we project the boundary states onto the ground states
of the free transfer matrix. In other words, the perturbation scheme makes the
Brioullin--Wigner
phase choice, in which the overlap of the perturbed (non-zero
gauge field) state with the unperturbed (zero gauge field)
state is real. Of course this is not the only way to fix the
ambiguity in the effective action, and in fact it is not an
adequate prescription for gauge fields with non-zero winding
number (instantons) for which this overlap is
zero\cite{Na:Ne:94b}: but it is a
perfectly adequate prescription for ordinary perturbation theory.

\section{The model, perturbation theory and the effective action}
\label{model}
The fermionic action in $d(=2n+1)$ dimensions proposed by Narayanan and
Neuberger
is\cite{Na:Ne:93,Na:Ne:94}
\be
S_{fermion}(\bar{\psi},\psi,U) = - \sum_{x,s,t} \bar{\psi}_s (x)
[\slash D + MP^R +
M^\dagger P^L ]_{st} \psi_t (x)
\label{Schiral}
\ee
where
\begin{eqnarray}
\slash D &=& \delta_{s,t} \frac{1}{2} \gamma_\mu (\nabla_\mu^\dagger +
\nabla_\mu )
 \nonumber \\
\nabla_\mu \psi_s (x) &=& U(x,\mu)\psi_s (x+\hat{\mu}) - \psi_s (x) \nonumber
\\
\nabla_\mu^\dagger \psi_s (x) &=& \psi_s (x) - U^{-1} (x-\hat{\mu},\mu)
\psi_s (x-\hat{\mu}) \nonumber
\\
M_{st} &=& \delta_{s+1,t} - \delta_{s,t} a_s \nonumber \\
M^\dagger_{st} &=& \delta_{s-1,t} - \delta_{s,t} a_s \nonumber \\
a_s &=& 1 - m_0 \mbox{sign} (s+\frac{1}{2}) - \frac{1}{2} \nabla^\dagger_\mu
\nabla_\mu \nonumber \\
P^R &=& \frac{1}{2} (1+ \gamma_d) \nonumber \\
P^L &=& \frac{1}{2} (1 - \gamma_d)
\label{Sfermion}
\end{eqnarray}
$\psi_s (x)$ and $\bar{\psi}_s(x)$ are Dirac spinors; $m_0$ is the domain wall
height, $0 < m_0 <1$, $x$ labels the sites on the $d-1$ dimensional ``real''
lattice, $s$ labels the infinite number of fermions, and as such can be seen as
a flavour index, or as the position variable in an ``extra'' $d$-th
dimension in
which the domain wall lives. We have set the lattice spacing $a$ (not to be
confused with $a_s$!) equal to one, but will restore an explicit $a$ to
expressions as we need to in taking the continuum limit ($a \rightarrow 0$)
later. The $\gamma_\mu, \mu=1,\cdots ,2n$ are Euclidean gamma
matrices. $P^R, P^L$ are the usual projection operators onto right and left
handed fermions respectively. Note that we take the gauge fields $U$ to be
$s$-independent, i.e. we have not introduced any extra degrees of
freedom for the gauge field.
The action above is explicitly invariant under both $s$--independent
local gauge transformations and global vector transformation.

More species\footnote{We are using
the word ``species'' rather than ``flavour'' simply because we
have already described the $s$--index on $\psi$ as a ``flavour''
index.} of fermion could be incorporated into
equation (\ref{Sfermion}) by simply adding more fermion fields $\psi$,
and changing the sign of the domain wall  mass $m_0$
according to whether a zero mode of a fermion field is
to be right-handed ($m_0 > 0$) or left-handed ($m_0 < 0$). That
is, we make a whole new copy of the action in equation
(\ref{Schiral}) for each new species of fermion.
The resulting action then has a  global ``vector'' invariance
$U(n_+) \otimes U(n_-)$, where $n_+$($n_-$) is the number of
fermion fields
with positive(negative) $m_0$.
This vector symmetry may be a candidate for a ``chiral'' symmetry
$U(n_+)_R\otimes U(n_-)_L$ for the zero modes ( there are $n_+$ right-handed
zero modes and $n_-$ left-handed zero modes by construction.).
For example a model with $n_+ = n_- = n_f$( as for QCD)
the symmetry becomes $U(n_f)_L \otimes
U(n_f)_R = U(1)_A \otimes U(1)_V \otimes SU(n_f)_A \otimes
SU(n_f)_V$.
The currents associated with this symmetry are
\begin{equation}
J_{\mu,a}^R = i\sum_{s=-\infty}^\infty \bar\psi^+_s \gamma_\mu T_a^R \psi^+_s,
\qquad J_{\mu,a}^L = i\sum_{s=-\infty}^\infty \bar\psi^-_s \gamma_\mu T_a^L
\psi_s^-
\end{equation}
where the $T_a^R$'s ($T_a^L$'s) are the generators of $U(n_+)$ ( $U(n_-)$ )
and the indices $\pm$ represent the sign of $m_0$.
(Here we omit the ``species'' index of $\psi^\pm$.)
These currents are not well-defined due to the infinite $s$
summation, so they will be redefined later in section
\ref{subsec}.

Equation (\ref{Sfermion}) is probably not the most familiar
way of writing
out the fermionic action for this model. For instance, we note that the lattice
derivative $\slash D$ is just the na{\"\i}ve derivative:
the Wilson terms appear in the
mass matrix $M$. To write down the action in a simpler fashion
(see reference \cite{Na:Ne:94} for instance),
we would start with the free fermion action with Wilson terms in
all $d$ dimensions, gauge $d-1$ of the $d$ dimensions and add a
domain wall mass in the $d$-th dimension.
In equations (\ref{Schiral}), (\ref{Sfermion}) the Wilson term for the
$d$-th dimension is obscured by
the
fact that the relevant $\gamma_d$ matrices are hidden in $P^L, P^R$.

The reason for writing the action in this way is that (\ref{Schiral}) is
clearly of the
general form (\ref{Lchiral}) first put forward in reference \cite{Na:Ne:93} as
a way of understanding different schemes\cite{Ka:92,Fr:Sl:93,Na:Ne:93} for
implementing chiral fermions. As we mentioned above,
these schemes all have in common the idea that in
order to create a chiral fermion (say right-handed), the mass matrix $M$ in
equation (\ref{Schiral}) should have a zero mode, while $M^\dagger$ should not.
This cannot be achieved with a finite dimensional $M$, so we must have an
infinite number of auxiliary fields. These fields may be Pauli-Villars
regulators, of alternating statistics\cite{Fr:Sl:93,AK,Fu:94} or fermion
fields labelled by $s$ and coupled to a domain
wall in this ``internal'' space, as in the scheme under investigation.
Of course these two approaches do not exhaust
the possibilities, but they are the only ones to have been investigated so far.

In order to do perturbation theory we need expressions for the fermion
propagator and
the vertices. The main complications arise from the rather messy form of the
propagator, first derived in reference \cite{Na:Ne:93}. Because translational
symmetry is broken in the extra dimension, we work in momentum space for the
$d-1$
dimensional ``real world'' and position space for the extra dimension. Then the
propagator is  (note that $\bar{p}_\mu = \sin (p_\mu a),
\hat{p}_\mu =
2 \sin (\frac{1}{2} p_\mu a)$):
\begin{eqnarray}
S_F (p) & = & ( -i \gamma_\mu \bar{p}_\mu + \frac{1}{2} \hat{p}^2 - M(p) P^R -
M^\dagger (p) P^L )^{-1} \nonumber \\
& = & \left( -i \gamma_\mu \bar{p}_\mu + M^\dagger (p) \right)P^R G^R
+ \left(-i \gamma_\mu \bar{p}_\mu + M(p) \right) P^L G^L \mbox{,}
\label{prop}
\end{eqnarray}
where
\begin{eqnarray}
M_{st} &=& \delta_{s+1,t} - \delta_{s,t} \tilde{a}_s (p) \nonumber \\
M^\dagger_{st} &=& \delta_{s-1,t} - \delta_{s,t} \tilde{a}_s (p) \nonumber \\
\tilde{a}_s (p) &=& \left\{ \begin{array}{ll} a_+ & s \ge 0 \\
                                  a_- & s < 0
\end{array} \right. \nonumber \\
a_\pm &=& 1+ \frac{\hat{p}^2}{2} \mp m_0 \mbox{,}
\end{eqnarray}
and
\begin{eqnarray}
G^L_{st} (p) &=& G^L_{ts} (p)  =
\left(\frac{1}{\bar{p}^2 + M^\dagger M} \right)_{st}
\nonumber \\
&=& \left\{ \begin{array}{ll}
Be^{-\alpha^+ |s-t|} + (A^L-B) e^{-\alpha^+ (s+t)} & s,t \ge 0 \\
A^L e^{-\alpha^+ s + \alpha ^- t} & s \ge 0,t < 0 \\
C e^{-\alpha^- |s-t|} + (A^L -C) e ^{\alpha^- (s+t)} & s,t < 0
\end{array} \right. \nonumber \\
G^R_{st} (p) &=& G^R_{ts} (p) =
\left(\frac{1}{\bar{p}^2 + MM^\dagger} \right)_{st} \nonumber \\
&=& \left\{ \begin{array}{ll}
B e^{-\alpha^+ |s-t|} + (A^R-B)e^{-\alpha^+ (s+t+2)} & s,t \ge -1 \\
A^R e^{-\alpha^+ (s+1) + \alpha^- (t+1)} & s \ge -1, t < -1 \\
C e^{-\alpha^- |s-t|} + (A^R - C) e^{\alpha^- (s+t+2)} & s,t < -1
\end{array} \right. \nonumber \\
\alpha^\pm &=& \cosh^{-1} \left[\frac{1}{2} \left(a^\pm +
\frac{1+\bar{p}^2}{a^\pm} \right) \right] \ge 0  \nonumber
\end{eqnarray}
\be
A^R = \frac{1}{a^- e^{\alpha^-} - a^+ e^{-\alpha^+}}, \hspace{3em}
A^L = \frac{1}{a^+ e^{\alpha^+} - a^- e^{-\alpha^-}} \nonumber
\ee
\be
B = \frac{1}{2a^+ \sinh \alpha^+}, \hspace{3em}
C = \frac{1}{2a^- \sinh \alpha^-} \mbox{.}
\ee
Note that the above form of the fermion propagator is valid only
for $s$--space infinite.

To obtain the vertices we introduce gauge fields $A_\mu$, defined by
\be
U(x, \mu) = e^{iea A_\mu (x)}
\ee
The vertices are somewhat simpler in form than the propagator:
in fact they are exactly the usual
Wilson vertices (see figure 1), obeying the lattice Ward
Identity:
\be
V^{(n)}_{\mu_1 \cdots \mu_n} (q, q') = \frac{a^{n-1} (-e)^n}{n !}
\sum_\mu \delta_{\mu \mu_1} \cdots \delta_{\mu \mu_n} \sum_s \delta_{s s_1}
\cdots \delta_{s s_n} \partial_\mu^{n} S_F^{-1} (\frac{q + q'}{2}).
\label{vertex}
\ee
where we have restored the dependence on the lattice spacing $a$ explicitly,
and $\partial_\mu^{n} S_F^{-1} (q)$ means $\partial^n S_F^{-1}
/\partial (q_\mu a)^n $. This is exactly the usual Wilson vertex, whose only
momentum dependence is through the sum of the ingoing and outgoing fermion
momenta, with a
trivial $s$-dependence added in. We note that there are an infinite number of
``seagull'' vertices, but with the addition of each photon the vertex decreases
by a factor of $a$. We will need only $V^{(1)}_\mu$ and $V^{(2)}_{\mu, \nu}$:
\begin{eqnarray}
V^{(1)}_\mu (q, q')  &=& (-e) \partial_\mu S_F^{-1} (\frac{q + q'}{2})
\nonumber \\
V^{(2)}_{\mu \nu} (q, q')  &=& a \frac{e^2}{2} \delta_{\mu \nu} \partial_\mu^2
S_F^{-1}
(\frac{q + q'}{2})
{}.
\label{v12}
\end{eqnarray}
We have left off the trivial $s$-dependence.

The bulk of this paper is devoted to the calculation of the
vacuum polarization tensor $\Pi_{\mu \nu} (p)$ (see figure 2)
for the chiral Schwinger model($d=3$) :
\be
\Pi_{\mu \nu} (p) = \Pi_{\mu \nu}^{(a)} (p) + \Pi_{\mu \nu}^{(b)} (p)
\ee
where $\Pi_{\mu \nu}^{(a)} (p)$ is the nonseagull diagram:
\be
\Pi^{(a)}_{\mu \nu} (p) = e^2 \int \frac{\mbox{d}^2 q}{(2 \pi)^2}
\sum_{st}
\mbox{Tr} \left\{ \partial^\mu S_F^{-1} (q) \left(S_F
(q-p/2) \right)_{st} \partial^\nu S_F^{-1} (q) \left(S_F (q+p/2) \right)_{ts}
\right\} a^2
\label{Pia}
\ee
and $\Pi_{\mu \nu}^{(b)} (p)$ is the seagull diagram:
\be
\Pi^{(b)}_{\mu \nu} (p) = e^2 \int \frac{\mbox{d}^2 q}{(2 \pi)^2}
\sum_{st}  -\delta_{st} \delta^{\mu \nu} \mbox{Tr} \left\{
(\partial^\mu)^2 S_F^{-1} (q) \left(S_F (q) \right)_{ss} \right\} a^2 \mbox{.}
\label{Pib}
\ee
In (\ref{Pia}) and (\ref{Pib}) we have used the vertex factors in (\ref{v12}).

The one-loop
effective action, to second order in the gauge fields (We note
that the effective action for an odd number of gauge fields
vanishes, by Furry's  theorem: see the appendix for details) is then given by
\be
S_{eff}^{(2)} = \frac{1}{2} \int \frac{\mbox{d}^2 p}{(2 \pi)^2} \;
 \tilde{A}_\mu (p) \tilde{A}_\nu (-p) \Pi_{\mu \nu} (p) ,
\label{Seff}
\ee
where $\tilde{A}_\mu (p)$ is the Fourier transform of the gauge field:
\be
A_\mu (x) = \int \frac{\mbox{d}^2 p}{(2 \pi)^2} \; e^{ip.(x + \frac{1}{2}
\hat{\mu} )} \tilde{A}_\mu (p),
\ee

It is convenient for later calculations to work in a chiral basis with
\begin{eqnarray}
\gamma_\mu &=& \left( \begin{array}{cc} 0 & \sigma_\mu \\
                                       \sigma^\dagger_\mu & 0 \end{array}
\right) \nonumber \\
\sigma_\mu & = & \left\{ \begin{array}{ll}
1 & \mu = 1 \\
-i & \mu = 2
\end{array} \right. \nonumber \\
\gamma_3 &=& -i \gamma_1 \gamma_2 = \left( \begin{array}{cc} 1 & 0 \\
                                      0 & -1 \end{array}
\right)
\end{eqnarray}
so that
 $\{
\gamma_1, \gamma_2, \gamma_3 \}$ are just the usual three Pauli matrices.

In this basis we have:
\begin{eqnarray}
S_F (p) &=& \left( \begin{array}{cc}
M^\dagger G^R & -i \sigma . \bar{p}\; G^L \\
-i \sigma^\dagger . \bar{p} \; G^R & M G^L
\end{array} \right) \nonumber \\
\partial_\mu S_F^{-1} (p) &=& i \gamma_\mu \tilde{p}_\mu + \bar{p}_\mu =
\left( \begin{array}{cc}
\bar{p}_\mu   & i \sigma_\mu \tilde{p}_\mu \\
i \sigma_\mu^\dagger \tilde{p}_\mu & \bar{p}_\mu
\end{array} \right) \nonumber \\
\partial_\mu^2 S_F^{-1} (p) &=& -i \gamma_\mu \bar{p}_\mu + \tilde{p}_\mu =
\left( \begin{array}{cc}
\tilde{p}_\mu & -i \sigma_\mu \bar{p}_\mu \\
-i \sigma_\mu^\dagger \bar{p}_\mu & \tilde{p}_\mu \mbox{.}
\end{array} \right)
\label{chiral}
\end{eqnarray}
Here $\tilde{p}_\mu = \cos (p_\mu a)$.

\section{Curing the $s$-divergences}
\label{subsec}
Because of the sums over infinite $s$-space
the effective action (\ref{Seff}) is ill-defined.
In order to make the effective action well-defined, we adopt the following
prescription. We
restrict the interaction between gauge
fields and fermions
to the range $ -L \le s \le L $, though the $s$-space itself
is infinite. We immediately see that the effective action (\ref{Seff})
diverges as $L\rightarrow \infty$.
The divergence, which arises from
the translationally invariant part of the fermion propagator, can be removed by
a subtraction
\be
S_{eff}^{tot} = S_{eff} - \frac{1}{2} (S_{eff}^+ + S_{eff}^- )
\mbox{,}
\label{sub}
\ee
where $S_{eff}^\pm$ arises from calculating (\ref{Seff}) with a constant mass
term $\pm m_0$. This is the prescription suggested in
\cite{Na:Ne:93,Na:Ne:94}. In the homogeneous effective actions $S_{eff}^\pm$
we also  restrict
the gauge fields to the finite range $-L \leq s \leq L$.
  After performing the subtraction in equation
(\ref{sub}) at finite $L$, we take the limit
$L\rightarrow \infty$.
Restricting the gauge fields to finite ranges
does not change the form of the free fermion propagator,
equations (\ref{prop}) and (5), but it breaks the gauge invariance of the
action
(\ref{Schiral}) under $s$-independent transformations.
Of course we know that we need to break the gauge invariance in
the imaginary part of the effective action in order to recover
the anomaly. Our prescription breaks gauge invariance in both
the real and imaginary parts of the effective action when $L$
are finite, but when we take $L \rightarrow \infty$ gauge
invariance is recovered in the real part and broken only in the
imaginary part. This is a key advantage of the KNN scheme.

The definition of the effective action above also suggests that
the currents associated with the global
``vector'' transformations in section \ref{model} should be
modified in the following way
\begin{equation}
J_\mu^R = i\sum_{s=-\infty}^{\infty}\bar\psi_s \gamma_\mu \psi_s
\rightarrow
J_\mu^R = i\sum_{s=-L}^{L}\bar\psi_s \gamma_\mu \psi_s ,
\end{equation}
so that the ``vector'' transformation generated by the modified currents
becomes
\begin{equation}
\psi_s \rightarrow \psi_s' = \left\{
\begin{array}{rc}
e^{i\theta}\psi_s  &  -L \le s \le L \\
\psi_s	            &   |s| \ge L
\end{array}
\right.  .
\end{equation}
It should be noted that the action is no longer invariant under this
modified transformation due to the presence of the terms
$\bar\psi_s(MP^R+M^\dagger P^L)_{st}\psi_t$, and, as seen in
section \ref{contsec} and section \ref{massgap}, this breaking of
the ``vector'' symmetry leads to
anomalies in the fermion number currents.

In this section we show explicitly that this subtraction scheme
renders the real part of the effective action
finite, but that the imaginary part of the action (which leads to the anomaly)
is finite without subtraction. The homogeneous action $S_{eff}^\pm$ is purely
real (as we show in the next section), so the imaginary part of the action is
unaffected by the subtractions. As such it is unambiguous, in apparent
contradiction with the overlap calculation of reference \cite{Na:Ne:94}. In
section \ref{pertphase} we address this apparent contradiction.

To see that the effective action is divergent in $s$ is very straightforward.
Looking at the expression for $\Pi_{\mu \nu}^{(a)}$ (equation (\ref{Pia})) for
instance, using the chiral basis (\ref{chiral}) for the propagators and
vertices, taking the Dirac trace and then summing over $s$ and $t$ gives terms
of the form
\be
\sum_{st} (O_1)_{st} (O_2^\prime)_{ts} \mbox{,}
\ee
where $O, O^\prime$ come from the set $\{G^L,G^R,MG^L,M^\dagger G^R \}$
and $\displaystyle \sum_{st}\equiv \sum_{s=-L}^{L}\sum_{t=-L}^{L}$. A
subscript ``1'' means ``evaluated at momentum $q+p/2$'', while subscript ``2''
means ``evaluated at $q-p/2$''. All such terms diverge: looking for instance at
$\sum_{st} (G^L_1)_{st} (G^L_2)_{ts} $ we find
\begin{eqnarray}
\sum_{st} (G^L_1)_{st} (G^L_2)_{ts} &=& B_1 B_2 \sum_{s,t=0}^{L}
e^{-\alpha_1^+ |s-t|} e^{-\alpha_2^+ |s-t|}  \\
& & + C_1 C_2 \sum_{s,t=1}^{L} e^{-\alpha_1^- |s-t|} e^{-\alpha_2^- |s-t|} +
\mbox{terms finite as $L \rightarrow \infty$} \nonumber
\label{GLGL}
\end{eqnarray}
We note that the divergence comes only from the translationally invariant part
of the propagator. As such it is natural to remove the divergence by
subtracting the effective action due to a homogeneous mass term. With a
constant mass term $\pm m_0$ instead of the domain wall we find that the
propagators are as in equation (\ref{prop}) above, but with $G^L = G^R = G^\pm,
M=M_\pm, M^\dagger = (M_\pm)^\dagger$, where
\begin{eqnarray}
G^+_{st} &=& B e^{-\alpha^+ |s-t|} \nonumber \\
G^-_{st} &=& C e^{-\alpha^- |s-t|} \nonumber \\
(M_\pm)_{st} &=& \delta_{s+1,t} - \delta_{s,t} a_\pm \nonumber \\
(M_\pm)_{st}^\dagger &=& \delta_{s-1,t} - \delta_{s,t} a_\pm \mbox{.}
\end{eqnarray}
{}From the form of the homogeneous propagator above, it is immediately obvious
that (\ref{sub}) is the correct prescription to render the effective action
finite. For instance, subtracting
\[
\frac{1}{2} \sum_{s,t=-L}^{L}
  (G^+_1)_{st}
(G^+_2)_{ts} + \frac{1}{2} \sum_{s,t=-L}^{L} (G^-_1)_{st} (G^-_2)_{ts}
\]
cancels the divergence in (\ref{GLGL}).

There are some more subtle points to be made: firstly, we note that only the
real part of the effective action is initially divergent: the
imaginary part is finite {\em before} subtractions. To
illustrate this we note that the ``$G^L G^L$'' term, for instance, is paired
with a ``$G^R G^R$'' term in the following way:
\be
\sum_{st} \left[ (G^L_1)_{st} (G^L_2)_{ts} \zeta_{\mu \nu}
+ (G^r_1)_{st} (G^R_2)_{ts} \zeta_{\mu \nu}^\star \right]
\mbox{,}
\ee
where
\be
\zeta = \sigma . \overline{q + p/2} \; \sigma . \overline{q - p/2} \;
 \sigma_\mu \sigma_\nu \tilde{p}_\mu
\tilde{p}_\nu
\mbox{.}
\ee
We can split the above sum into its real and imaginary
parts as follows:
\begin{eqnarray}
\lefteqn{\sum_{st} \left[ \left( (G_1^L)_{st} (G^L_2)_{ts} + (G_1^R)_{st}
(G_2^R)_{ts}
\right) \left( \frac{\zeta_{\mu \nu} + \zeta_{\mu \nu}^\star}{2} \right)
\right. } \nonumber \\
& & \left. + \left( (G_1^L)_{st} (G_2^L)_{ts} - (G_1^R)_{st}(G_2^R)_{ts}
\right)
\left( \frac{\zeta_{\mu \nu} - \zeta^\star_{\mu \nu} }{2} \right) \right]
\mbox{.}
\label{GLGLGRGR}
\end{eqnarray}
It is easily shown that $\sum_{st} (G_1^R)_{st} (G_2^R)_{ts}$ diverges in
exactly the same way as $\sum_{st} (G_1^L)_{st} (G_2^L)_{ts}$, so that the
first term in (\ref{GLGLGRGR}) is infinite, but the second is finite. In the
next section we show that the homogeneous actions $S_{eff}^\pm$ are purely
real, so that the imaginary part of the effective action is unaffected by the
subtractions, and is hence unambiguous.

\section{Reality of the homogeneous effective action}
\label{homreal}
This can be seen by a brief but sloppy argument, or a slightly longer explicit
calculation. The sloppy argument first: given that we have fermions coupled to
an Abelian gauge
field in three dimensions, we might expect an imaginary piece in
the homogeneous effective action, of Chern-Simons form\cite{Co:Lu:89,So}:
\be
\int \mbox{d}^3 x \epsilon_{\alpha \beta \gamma} A_\alpha (x) \partial_\beta
A_\gamma (x),
\label{CS}
\ee
where $\partial_\beta$ here means $\partial / \partial x_\beta $. Now remember
that our gauge
field has only two components, so the integrand in equation (\ref{CS}) above
reduces to
\be
A_2(x) \partial_3 A_1 (x) - A_1 (x) \partial_3 A_2 (x) ,
\ee
where $\partial_3 = \partial / \partial x_3 = \partial / \partial s$. If the
gauge fields are $s$-independent as in the KNN case, then the Chern Simons term
vanishes. Of course this does not rule out other imaginary terms, so we really
should do an explicit calculation.

Looking first at the seagull contribution to $S_{eff}^\pm$, with the two gauge
field
vertex, we find that it is proportional to the integral over $q$ of
\begin{eqnarray}
\lefteqn{
\sum_s  \mbox{Tr} \left[ \left(
\begin{array}{cc} (M^\dagger G) & -i \sigma . \bar{q} G \\
                    -i \sigma^\dagger . \bar{q} G    & M G
\end{array}
\right)_{ss}
\left(
\begin{array}{cc} \tilde{q}_\mu & -i \sigma_\mu^\dagger \bar{q}_\mu  \\
                  -i \sigma_\mu \bar{q}_\mu   & \tilde{q}_\mu
\end{array}
\right)
\right]} \label{Jhom}  \\
&=& \sum_s  \left[(M^\dagger G)_{ss} \tilde{q}_\mu
+ (M G )_{ss} \tilde{q}_\mu - G_{ss} \sigma_\mu \sigma . \bar{q}
\bar{q}_\mu
-G_{ss} \sigma_\mu^\dagger \sigma^\dagger . \bar{q} \bar{q}_\mu \right]\mbox{.}
 \nonumber
\end{eqnarray}
For simplicity we have left off the ``$\pm$'' sub- and superscripts.
This sum is divergent, but real: we get no
contribution to the imaginary part of $S_{eff}^\pm$.

To see that the non-seagull term is real is a bit (but not much) more subtle.
Writing
out the vertices and propagators in the chiral basis as in equation(\ref{Jhom})
above we obtain
\begin{eqnarray}
\lefteqn{ \Pi_{\mu \nu}^{(a)} (p) } \nonumber \\
&=& e^2 \int \frac{\mbox{d}^2 q}{(2 \pi)^2}
 \sum_{st}  \left[ G_1 (M^\dagger G)_2 (\zeta_{\mu
\nu})_1^\star + G_1 (M G)_2 (\zeta_{\mu \nu})_1 \right. \nonumber \\
& & \left. + (M^\dagger G)_1 G_2 (\zeta_{\mu
\nu})_2 + (M G)_1 G_2 (\zeta_{\mu \nu})_2^\star + \mbox{other terms}
\right]_{st}
\label{Ihom}
\end{eqnarray}
where
\begin{eqnarray}
(\zeta_{\mu \nu})_1 & = & \sigma^\dagger . \overline{q + p/2} \;
 \sigma_\mu^\dagger \tilde{q}_\mu \bar{q}_\mu
\nonumber \\
(\zeta_{\mu \nu})_2 & = & \sigma^\dagger . \overline{q - p/2} \;
 \sigma_\mu^\dagger \tilde{q}_\mu \bar{q}_\mu
\mbox{.}
\end{eqnarray}
The subscripts ``1'' and ``2'' on $G$, $MG$ or $M^\dagger G$ again mean
 ``evaluated at
momentum $q + p/2$'' or ``evaluated at momentum $q-p/2$''respectively.
The imaginary part of the expression in (\ref{Ihom}) is
\begin{eqnarray}
\lefteqn{ e^2 \int \frac{\mbox{d}^2 q}{(2 \pi)^2}
\sum_{st} \left[ \left( G_1 (M^\dagger G)_2 - G_1
(MG)_2 \right)  \left( \frac{(\zeta_{\mu
\nu})_1^\star -(\zeta_{\mu \nu})_1)}{2i} \right)  \right.}\nonumber  \\
& & \left. -\left( G_2 (M^\dagger G)_1  - G_2 (MG)_1 \right)
\left( \frac{(\zeta_{\mu
\nu})_2^\star -  (\zeta_{\mu \nu})_2}{2i} \right) + \mbox{other terms}
\right]_{st}
\mbox{.}
\label{Ihomim}
\end{eqnarray}
This sum can be easily shown to converge (the subtractions ensure this). Thus
we are justified in doing the $q$ integral {\em inside} the sum. Putting $q
\rightarrow -q$ in the
second term in the square brackets above
 turns all the subscript 1's into 2's and vice versa: for $G$,
$MG$ and $M^\dagger G$ this is obvious since these are only functions of $q$
through $|q \pm
p/2|$. The $\zeta$ terms are also easily seen to interchange the ``1''
and ``2'' subscripts. Then the first term in equation (\ref{Ihomim})
cancels with the second and we get
zero for the imaginary part once more.

In equation (\ref{Ihom}) we only listed four of sixteen terms, but the
argument goes through in a similar fashion for all the others.

The last two sections have shown that the imaginary part of the effective
action is finite and unambiguous, in apparent contradiction with the
calculation of reference \cite{Na:Ne:94}. In the following section we look at
this apparent contradiction.

\section{Perturbation theory and the phase ambiguity}
\label{pertphase}
The overlap formula\cite{Na:Ne:94} for the effective action in our scheme
is of the form
\begin{eqnarray}
\exp [ S_{eff} (A_\mu)] & = & \lim_{L\rightarrow \infty}
[ \lim_{s_0 \rightarrow \infty} \langle b - | (\hat{T}_- (0))^{s_0}
\nonumber \\
&\times & (\hat{T}_- (A_\mu))^{L}
(\hat{T}_+ (A_\mu))^{L} (\hat{T}_+ (0))^{s_0}
| b + \rangle ]  \nonumber \\
 & = & \langle b - | 0 -\rangle \langle 0 + | b + \rangle
\nonumber \\
&\times & \lim_{L\rightarrow \infty}
\langle 0 - |
(\hat{T}_- (A_\mu))^{L}
(\hat{T}_+ (A_\mu))^{L}
| 0 + \rangle   ,
\label{overlap}
\end{eqnarray}
where $\hat{T}_\pm (A_\mu)$ are the transfer matrices and
$|0 \pm \rangle$ are the ground states of the free transfer
matrices. The point is that when we take $s_0 \rightarrow \infty$
we project the boundary states $| b \pm \rangle$  onto the
ground states of the free transfer matrix.
Since $| b \pm \rangle $ is naturally taken to be independent of $A_\mu$,
$\langle b - | 0 - \rangle \langle 0 + | b + \rangle$ does not depends
on $A_\mu$ at all, and as such it affects only the irrelevant
{\it constant} part of the effective action.
We can see also that the
boundary conditions for the fermions do not affect the final
result.
Even periodic boundary conditions for fermions give the same result
as long as the limit $s_0\rightarrow \infty$ is taken {\it before}
taking the limit $L\rightarrow \infty$. (To get a non-zero result the condition
$\langle 0 + | 0 - \rangle \not= 0$ is also needed.)
Besides the irrelevant constant
$\langle b - | 0 - \rangle \langle 0 + | b + \rangle$  the final
answer is the same answer we would have got if we had taken $|b \pm \rangle = |
0 \pm \rangle$ in the first place, which is the Wigner-Brioullin phase
choice\cite{Na:Ne:93b,Na:Ne:94},
where the overlap of the
perturbed state $\lim_{L \rightarrow \infty}[\lim_{s_0
\rightarrow \infty} (\hat{T}_\pm (A_\mu))^{L} (\hat{T}_\pm
(0))^{s_0} ] | b \pm \rangle $
with the unperturbed state
$\lim_{s_0, L \rightarrow \infty} (\hat{T}_\pm
(0))^{s_0+L} | b \pm \rangle $ is real.

\section{The effective action in the continuum limit}
\label{contsec}
In section \ref{subsec} we showed that the subtractions render the $s,t$-sums
in the
effective action in (\ref{Seff}) finite. But we still have to integrate over
$p$ and $q$. It can easily be checked that after the subtractions the integrand
has no singularities in $p$ or $q$ when we take the continuum limit, $a
\rightarrow 0$, other than the singularity noted in reference \cite{Na:Ne:93}:
\be
\lim_{a \rightarrow 0} A^L (p) = \frac{m_0 (4-m_0^2)}{4 p^2 a^2} \mbox{.}
\ee
This part of the
fermionic propagator corresponds to the zero mode bound to the
domain wall.
The zero mode is absent in the homogeneous propagators (there is no domain wall
for it to be bound to), and in fact the homogeneous action will give no
contribution in
our continuum calculation. We will start by just leaving the homogeneous terms
out entirely, but justify our rashness explicitly as we go along.
The homogeneous action has of course already
fulfilled its role, to tame the $s$-divergence of the integrand so that
meaningful statements can be made about its continuum limit. We
note that the method used in this section is basically identical
to that used in reference \cite{Ao:Hi:93} on Kaplan and Shamir
fermions. We explicitly use Karsten and Smit's
approach\cite{Ka:Sm:81} to momentum integration on the lattice.

We wish to evaluate (\ref{Seff}) in the continuum limit, $a \rightarrow 0$. We
will see that because of the divergence of the propagators at small momentum
it is natural to divide the region of $q$-integration, $A = \{
(q_1,q_2) : |q_1| < \pi /a, |q_2| < \pi /a \} $ into an ``inside'',
$A_1 = \{
(q_1,q_2) : |q_1| < \epsilon /a, |q_2| < \epsilon /a \} $ and an ``outside'',
$A_2 = A - A_1$ (see reference \cite{Ka:Sm:81}, p 121-122).
$\epsilon$ is a small positive number which we take to zero only {\em after} we
take $a \rightarrow 0$. In the outside region $A_2$ we can rescale $q
\rightarrow q^\prime
= qa$ and take $a \rightarrow 0$ in the integrand.

However in the inside region $A_1$ we cannot do this asymmetric rescaling,
since we do not
have a guarantee that $q^\prime > \epsilon \gg pa$. In this region we must take
the $a \rightarrow 0$ limit of the integrand symmetrically. We get the
following
contribution to $\Pi_{\mu \nu}^{(a)} (p)$:
\begin{eqnarray}
\lefteqn{ e^2 \int_{A_1} \frac{\mbox{d}^2 q}{(2 \pi)^2}
 \sum_{st} \mbox{Tr} [i \gamma^\mu \left( -i
\gamma^\alpha (q-p/2)_\alpha a \right) G_0^L (q-p/2) P^L } \\
& & \times i \gamma^\nu (-i \gamma^\beta (q+p/2)_\beta a) G_0^L (q+p/2)P^L ]a^2
\mbox{,}
\nonumber
\end{eqnarray}
where
\be
G_0^L (q)_{st} = \lim_{a \rightarrow 0} G^L (q)_{st} = \frac{1}{q^2 a^2} F^L
(s,t)
\label{G0L}
\ee
and
\be
F^L (s,t) = F^L (t,s) = \frac{m_0 (4-m_0^2 )}{4} \times
\left\{ \begin{array}{ll} (1-m_0)^{s+t} & s,t \ge 0 \\
                          (1-m_0)^s (1+m_0)^t & s \ge 0, t < 0 \\
                          (1+m_0)^{s+t} & s,t <0
\end{array} \right.
\ee
The
$M$ terms in the propagators give zero because of the Dirac trace.
We have taken the $a \rightarrow 0$ limit of the integrand.
Note that we have interchanged the order of the limit and the sum, which is
valid only because we have explicitly shown that the $s,t$-sum is finite. The
continuum limit sum has been done in reference \cite{Ao:Hi:93} and is a very
simple result:
\be
\sum_{st} F^L (s,t)^2 = \sum_{s} F^L (s,s) = 1 \mbox{.}
\label{FLsq}
\ee
The contribution of the seagull graph $\Pi_{\mu \nu}^{(b)}$ to the inner region
is of order $a^2$, as are the contributions from the
homogeneous effective actions. In the latter case, though, we note (at the risk
of being suffocatingly pedantic) that the subtractions {\em were} needed to
make the $s,t$-sums finite first, so that the continuum limit made sense.

So now all that remains to be done is the $q$ integral:
\be
\int \frac{\mbox{d}^2 q}{(2 \pi)^2} \; \mbox{Tr} (P^L \gamma^\mu \gamma^\alpha
\gamma^\nu \gamma^\beta ) \frac{ (q-p/2)_\alpha (q+p/2)_\beta }{(q-p/2)^2
(q+p/2)^2} \mbox{.}
\ee
The two-dimensional Dirac trace is
\begin{eqnarray}
\lefteqn{\mbox{Tr} (P^L \gamma^\mu \gamma^\alpha
\gamma^\nu \gamma^\beta ) =} \\
& &  g^{\mu \alpha} g^{\nu \beta} - g^{\mu \nu}
g^{\alpha \beta} + g^{\mu \beta} g^{\alpha \nu} -i \delta^{\mu \alpha}
\epsilon^{\nu \beta} -i \delta^{\nu \beta} \epsilon^{\mu \alpha} \mbox{.}
\end{eqnarray}
Note that if we replaced $P^L$ by $P^R$ in the above equation (as we would do
to make a left-handed fermion), the imaginary part of the trace would reverse
sign: this will also lead to an anomaly with the opposite sign.
The integral at first sight looks logarithmically divergent, but in fact it
converges, so we can let the integration region run from
$-\infty$ to $\infty$ (remember $\epsilon$ is
to
be taken to zero only after $a \rightarrow 0$)
and the integral gives (using either Feynmann parameters or exponentiation of
the denominators):
\be
\int \frac{\mbox{d}^2 q}{(2 \pi)^2} \; \frac{(q-p/2)_\alpha (q+p/2)_\beta}
{(q-p/2)^2 (q+p/2)^2}
= - \frac{1}{4 \pi} \frac{p_\alpha p_\beta}{p^2} \mbox{,}
\ee
so the final contribution to $\Pi_{\mu \nu} (p)$ from the inner region is
\be
 \frac{e^2}{4 \pi} \frac{1}{p^2}
\left[
i \epsilon^{\mu \alpha} p_\alpha p_\nu + i \epsilon^{\nu \alpha} p_\alpha p_\mu
+ 2(\delta_{\mu \nu} p^2 - p_\mu p_\nu ) -  \delta_{\mu \nu} p^2
\right] \mbox{.}
\label{Pi1}
\ee
The very last term in (\ref{Pi1}) makes us feel slightly uncomfortable because
it breaks gauge invariance (in the
two-dimensional sense) but is not the
usual anomaly because it is real (the anomaly terms are in fact the first two
terms in (\ref{Pi1})). Such a ``longitudinal'' term
was found for models with $s$-dependent gauge fields\cite{Ao:Hi:93}.
Thankfully the contribution from the outside region
 exactly cancels this term.

The contribution to $\Pi_{\mu \nu}$ from the outside region of
integration $A_2$ is as follows:
\be
 e^2  \int_{A_2^\prime}
 \frac{\mbox{d}^2
q}{(2 \pi)^2} \; \sum_s \left[
\mbox{Tr} \{ \partial_\mu S_F^{-1} (q) \partial_\nu S_F
(q) \}_{ss} + \delta^{\mu \nu} \mbox{Tr} \{ \partial_\mu^2 S_F^{-1} .
S_F \}_{ss}
\right] \mbox{,}
\label{Pi2}
\ee
where in the first term we have used $\partial_\nu S_F^{-1} = -S_F^{-1}
\partial_\nu S_F . S_F^{-1} $. We have done the rescaling $q \rightarrow
q^\prime = qa$ and dropped the primes. The integration region $A_2^\prime$ is
given by $A_2^{(a)} \cup A_2^{(b)} \cup A_2^{(c)} \cup A_2^{(d)}$, where
\begin{eqnarray}
A_2^{(a)} &=& \{ (q_1,q_2) : q_1 \in (\epsilon, \pi), q_2 \in (-\pi, \pi) \}
\nonumber \\
A_2^{(b)} &=& \{ (q_1,q_2) : q_1 \in (-\pi,-\epsilon), q_2 \in (-\pi, \pi) \}
\nonumber \\
A_2^{(c)} &=& \{ (q_1,q_2) : q_1 \in (-\epsilon, \epsilon), q_2 \in (\epsilon,
\pi)
\}
\nonumber \\
A_2^{(d)} &=& \{ (q_1,q_2) : q_1 \in (-\epsilon, \epsilon), q_2 \in (-\pi,
-\epsilon)
\}   \mbox{.}
\label{A2prime}
\end{eqnarray}
Let us first look at the case $\mu = \nu$. Then we note that
\be
\partial_\mu S_F^{-1} \partial_\mu S_F + \partial_\mu^2 S_F^{-1} . S_F
= \partial_\mu (\partial_\mu S_F^{-1} . S_F) \mbox{,}
\ee
and putting $\mu = 2$ for definiteness, we have the following integral to
evaluate:
\be
\int \frac{\mbox{d} q_1}{(2 \pi)^2} \; \sum_s \left[ \mbox{Tr} \left\{
\frac{\partial}{\partial q_2} S_F^{-1} . S_F \right\}_{ss} \right]_{q_2 =
(q_2)_{min}}^{q_2 =
(q_2)_{max}} \mbox{.}
\label{Pi2int}
\ee
We note that the integration region $A_2^\prime$ involves large momenta (our
rescaled $q$ of order $1$), so at this point we have no justification to
replace the propagators in (\ref{Pi2int}) with their zero mode piece. However
noting that the terms in square brackets in (\ref{Pi2int}) must be odd in
$q_2$ to contribute, we find that all such terms have a factor of $\bar{q}_2
= \sin q_2$ (remember we have rescaled $q$ so it is as if $a=1$), and as such
give zero at the integration limits in (\ref{A2prime}) where $p_2 = \pi$ or
$-\pi$. In particular, regions $A_2^{(a)}$ and $A_2^{(b)}$ give zero, and
$A_2^{(c)}$ and $A_2^{(d)}$ combine to give
\be
- \int_{-\epsilon}^\epsilon \frac{\mbox{d} q_1}{(2 \pi)^2} \; \sum_s
\left[ \mbox{Tr}
\left\{ \gamma_2^2 \tilde{q}_2 \bar{q}_2 (P^R G^R + P^L G^L)_{ss} +
\bar{q}_2 (M^\dagger G^R P^R + M G^L P^L)_{ss} \right\} \right]_{q_2 =
- \epsilon}^{q_2 = \epsilon}
\ee
Now that all momenta are small ($< \epsilon$) we see that only the zero
mode part of the propagator in the $G^L$ term contributes
(the other terms are down by a factor
of $\epsilon$, as indeed are the terms that would have come from the
homogeneous effective action), and we get (noting that $\sum_s F^L (s,s) = 1$
\cite{Ao:Hi:93}):
\begin{eqnarray}
\lefteqn{2 \mbox{Tr} (\gamma_2^2 P^L)
\int_{-\epsilon}^\epsilon \frac{\mbox{d} q_1}{(2 \pi)^2} \;
\frac{\epsilon}{q_1^2 + \epsilon^2} } \nonumber \\
& & = \frac{1}{2 \pi^2} \left[ \tan^{-1} \frac{q_1}{\epsilon} \right]_{q_1 =
-\epsilon}^{q_1 = \epsilon} = \frac{1}{4 \pi}.
\end{eqnarray}
Applying similar reasoning to the cases $\mu =1$, and to $\mu \ne \nu$, we
finally obtain for the integral in (\ref{Pi2})
\be
\frac{e^2}{4 \pi} ( \delta_{\mu \nu} + i \epsilon_{\mu \nu} )
\ee
The first term is exactly what was needed to cancel the last term in
(\ref{Pi1}), as advertised; the second term would give rise to a Chern-Simons
interaction in the effective action (\ref{Seff})
if the gauge fields were $s$-dependent, but with $s$-independent
gauge fields gives no contribution to the effective action. We
omit this term from now on.

The final result for the continuum limit of the vacuum polarization is then
\be
\Pi_{\mu \nu} (p) =
 \frac{e^2}{4 \pi} \frac{1}{p^2}
\left[
i \epsilon^{\mu \alpha} p_\alpha p_\nu + i \epsilon^{\nu \alpha} p_\alpha p_\mu
+ 2(\delta_{\mu \nu} p^2 - p_\mu p_\nu )
\right] \mbox{.}
\label{Picont}
\ee
The one-loop effective action is then given by (\ref{Seff}). The consistent
anomaly
${\cal A} (x)$ is defined as the variation of the effective action under a
gauge transformation $A_\mu \rightarrow A_\mu + \partial_\mu \Lambda$:
\be
{\cal A} (x) = -  \frac{\delta S_{eff}}{\delta \Lambda} =
e \partial_\mu J_\mu (x),
\ee
where $J_\mu (x)$
is the fermion number current
\be
J_\mu (x) = \frac{1}{e} \frac{\delta S_{eff} }{\delta A_\mu (x) }.
\label{current}
\ee
In momentum space we have
\be
{\cal A}_p = p_\mu \tilde{J}_\mu (p) =
i \frac{e^2}{4 \pi} p_\mu \epsilon^{\mu \nu} \tilde{A}_\nu (p).
\label{anomaly}
\ee
We make note of two points: firstly we have calculated the
consistent form of
the anomaly; the covariant form for the Abelian theory differs only by a factor
of $2$ (i.e. the factor $4 \pi$ in the denominator is replaced by $2 \pi$).
Secondly, we should discuss how to make an anomaly-free theory. We
have already checked that for a fermion of opposite chirality, the
anomaly reverses sign, so the vector theory with a right-handed and a
left-handed fermion is anomaly-free. We could also
implement the $3-4-5$ model, with say two
right-handed particles with charges
$e^R_1 =3e$, $e^R_2 =4e$ and one left-handed particle
with charge $e^L_5 = 5e$:
it is clear from the above arguments that the anomaly
vanishes in the continuum and we can also implement this model on the lattice.

\section{Graphs with more than two external photons}
\label{higherorder}
In this section we show that graphs having fermion loops with more than two
photons attached
 vanish in the continuum limit (as they do in the continuum theory, this fact
making the model easily solvable).
This will allow us to give exact results for the mass
gap and the chiral order parameter in the continuum limit, for the vector-like
theory (the usual Schwinger model). We must first show
that the higher order graphs are $s$-finite, then look at the momentum
integrals in the continuum limit, using the same division into ``inside'' and
``outside'' regions that we used in section \ref{contsec}.

\subsection{$s$-subtractions}
The $s$-subtractions render graphs with more than two external photon lines
$s$-finite because, as for the
vacuum polarization graph,
the $s$-divergence comes from the translationally invariant
parts of the propagators, and is exactly cancelled by the homogeneous
subtractions. Let us look for example at the $n^{th}$ order graph in figure
3. It is clear that the {\em most} $s$-divergent terms are of the
form
\be
\sum_{s_1, s_2, \cdots , s_n} e^{-\alpha_1 | s_1-s_2 |} e^{-\alpha_2 |s_2-s_3|}
\cdots e^{-\alpha_n |s_n - s_1|} .
\label{higherssum}
\ee
This divergence is exactly the one that will be cancelled by the corresponding
homogeneous terms. The only potential problem occurs if we take one of the
factors $\exp \{ - \alpha_i |s_i - s_{i+1} | \} $, and replace it with a
{\em less} dangerous part of the propagator, of form $\exp \{
-\beta_i s_i - \beta_{i+1} s_{i+1} \} $ (for $s_i, s_{i+1} \ge 0$). There is no
corresponding term from the homogeneous effective actions to cancel this term,
so it must be finite by itself. It is not immediately obvious that this is so,
given the $n-1$ dangerous-looking factors that are
left. However it is easy to see that in fact with this modification the sum in
equation (\ref{higherssum}) is finite. Consider making exactly the replacement
described above, and summing over $s_i$. It is easily shown that
\be
\sum_{s_i = 0}^{L} e^{-\alpha_{i-1} | s_{i-1}-s_i |} e^{-\beta_i s_i -
\beta_{i+1} s_{i+1} }
\ee
gives
factors of the form $\exp \{ -\beta_{i-1} s_{i-1} - \beta_{i+1} s_{i+1} \} $
or $\exp \{ -\beta_{i-1}' L - \beta_{i+1} s_{i+1} \}$.
But these factors are exactly what are needed to make
the sums over $s_{i-1}$ and
$s_{i+1}$ converge. Carrying out the sums over $s_1, \cdots s_n$, we arrive at
a finite answer. Replacing more than one of the factors in (\ref{higherssum})
in this way just makes the sum even more convergent, so we are done.

We note that the above argument still holds if we replace some of
the vertices in figure 3 with seagull vertices, because the
$s$-divergences come from the propagators: the $s$-dependence of
the vertices (equation (\ref{vertex})) is trivial.

\subsection{Integration over the ``inside'' region}
Having shown $s$-convergence, we can now worry about doing the momentum
integration in the continuum limit. In the ``inside'' region defined in section
\ref{contsec}, we can replace the propagators and vertices by their $a
\rightarrow
0 $  limits. We find immediately that (as for the vacuum polarization)
any graph
with a seagull vertex is down by factors of $a$ and gives vanishing
contribution to the ``inside'' integration. For the non-seagull graphs (see
figure 3) we
replace the propagators with their $a \rightarrow 0$ limit:
\be
\lim_{a \rightarrow 0} (S_F (p))_{st} = -i \gamma^\mu p_\mu a G_0^L (p) P^L ,
\ee
where $G_0^L (p)_{st}$ is given by equation (\ref{G0L}). We have shown that the
$s$-sum associated with figure 3 is finite but it still must be
done: however it is easily shown that
\be
\sum_{s_2 = -\infty}^\infty F^L (s_1,s_2) F^L (s_2,s_3) = F^L (s_1,s_3) ,
\label{FLFL}
\ee
so that
\begin{eqnarray}
\lefteqn{\sum_{s_1, s_2, \cdots , s_n} F^L (s_1,s_2) F^L (s_2, s_3)
\cdots F^L (s_n, s_1)} \nonumber  \\
& = & \sum_{s_n} F^L (s_n, s_n) \nonumber \\
& = & 1 ,
\end{eqnarray}
where the first equality follows from repeated  application of (\ref{FLFL}) and
the second from (\ref{FLsq}).

We note that because the $s$-sums just give a factor of unity, the propagators
and vertices may be replaced by the following $s$-independent forms and the
$s$-sums ignored:
\begin{eqnarray}
S_F (p) &=& \frac {1} {\slash p} P^L \nonumber \\
V_\mu &=& (-e) \gamma_\mu .
\label{contpropvert}
\end{eqnarray}
We can now follow the methods of reference \cite{Bo:Ko:87} to show that for $n
> 2$ the graphs of figure 3 vanish in the inner region. The
trick is to make use of the following vector and axial-vector Ward identities,
which hold for the continuum forms of the propagator and vertex in equation
(\ref{contpropvert}):
\begin{eqnarray}
S_F (p+l) l_\mu V_\mu S_F (p)& =& (-e) [ S_F (p) - S_F (p+l) ] \nonumber \\
S_F (p+l) l_\mu V_\mu^5 S_F (p)& =& (-e) [ \gamma_5 S_F (p) - \gamma_5
S_F (p+l) ]
\label{Ward}
\end{eqnarray}
(note that $V^5_\mu = (-e) \gamma_\mu \gamma_5 $). In two dimensions we have
the relationship $\gamma_\mu \gamma_5 = -i \epsilon^{\mu \nu} \gamma_\nu $, so
that we can relate the graph in figure 3 to the graph with the
vertex factors $\gamma_{\mu_i}$ replaced by $\gamma_{\mu_i} \gamma_5$. To carry
out the proof (whose details we do not repeat), one considers the subset of
graphs obtained by leaving the order of vertices $2, \cdots , n$ fixed but
attaching photon $1$ in any position relative to the other vertices. Using
(\ref{Ward}) one can show that the contraction of the sum of this subset of
graphs with
the external momentum $l_1$ of photon $1$ vanishes. This means that the sum of
this subset
of graphs has zero divergence. Carrying out the same procedure for the axial
graphs shows that the sum has zero curl as well,
meaning that it must be identically
zero.

But what about $n=2$, which we have already seen to give a finite
answer? Well of course we have been a bit sloppy: the above
argument only holds for $n > 2$, because for $n = 2$, the
contraction of the graph with an external momentum gives an
expression which is linearly divergent, and we are not allowed to
use the Ward identities in (\ref{Ward}).

\subsection{Integration over the ``outside'' region}
The integrals over the outside region vanish for $n > 2$, by a
simple power counting argument. In reference
\cite{Bo:Ko:87} it is shown that for $n > 2$ any graph with $n$
photons attached to a fermion loop (i.e. the graph in figure 3,
or any variation with seagull vertices) vanishes as $a
\rightarrow 0$, unless the propagator has a pole. Since in the
outside region we have
excluded the only pole in the propagator by cutting out the
region $|q_{\mu}| < \epsilon$, where $\epsilon$ was to be taken
to zero only {\em after} $a \rightarrow 0$, we are done.

\section{Comparison with other regularizations}
\label{massgap}
The vacuum polarization in the chiral Schwinger model has been
calculated by several authors, using both continuum
\cite{Ja:Ra:85,Ha:Ts:87} and lattice regularizations
\cite{Ao:88,Fu:Ka:88,Ki:et:al:88}. Their results may be
summarized by
\be
\Pi_{\mu \nu}^{\mbox{\small others}} (p) =
\Pi_{\mu \nu}^{\mbox{\small our}} (p) + \frac{e^2}{4 \pi} C
\delta_{\mu \nu}
\label{C}
\ee
where $\Pi_{\mu \nu}^{\mbox{\small our}} (p)$ is given by
equation (\ref{Picont}).
Here $C$ is a constant called the
regularization constant, which was allowed but undetermined in the continuum
calculations\cite{Ja:Ra:85,Ha:Ts:87}, and explicitly given in the
lattice calculations\cite{Ao:88,Fu:Ka:88,Ki:et:al:88}. In these
previous lattice regularizations, $C$ was non-zero and real,
and depended on the Wilson
parameter $r$. It emerges as a necessary consequence of the
Wilson formulation for removing the doubler modes. The problem
with a non-zero $C$ in equation (\ref{C}) is that this term
breaks gauge invariance in the
{\em real} part of the effective action, meaning that
the effective action for a left-handed fermion is {\em
not}  the complex conjugate of that for a right-handed
fermion.
In other words, to restore gauge invariance by say making a
vector theory with a left and a right-handed particle, we have to
give up the property that chiral determinants for left and right
handed particles are complex conjugates. A further peculiarity
relating to non-zero $C$ is that the chiral Schwinger model
develops a boson excitation of mass $e^2 (C+1)^2/4 \pi C$.
 So perhaps our  most important result is that our
$C$ is zero. The gauge boson
for the chiral Schwinger model then becomes
 infinitely heavy and decouples from the theory.
A gauge-invariant vector theory is easily
constructed by simply adding the effective actions for a
left-handed fermion and a right handed fermion. The vacuum
polarization  in the vector theory then changes from the
expression (\ref{Picont}) to
\be
\Pi_{\mu \nu}^{(L+R)} (p) = \frac{e^2}{\pi} \left(
\delta_{\mu \nu} - \frac{p_\mu
p_\nu }{p^2} \right)
\ee
Because graphs with fermion loops having more than two photons attached vanish,
we can obtain the exact current-current correlation function (and hence the
mass gap)
by just summing bubble graphs, exactly as in the continuum
theory.  We simply quote the result from reference
\cite{Bo:Ko:87}, noting that it is exactly the result of the continuum theory:
\be
\mu = \frac{e}{\sqrt{\pi}}.
\ee
We can also get a result for the chiral order parameter $\langle \bar{\psi}
\psi \rangle$. This is zero in the perturbative vacuum, but may still be
calculated perturbatively from the four-point function. Once again we simply
quote the result, referring the interested reader to reference \cite{Bo:Ko:87}
and the references therein:
\be
\langle \bar{\psi} \psi \rangle = \frac{\mu^2}{4 \pi^2} e^{2 \gamma_E},
\ee
where $\gamma_E$ is Euler's constant. The actual results are not terribly
important for our purposes: our main purpose in quoting them is to emphasise
that we have obtained the correct continuum limit at all orders in perturbation
theory. Of course the main result we needed was that, as for the continuum
theory, fermion loops with more than two photons attached vanish, so that
perturbation theory is rather easily summed.

We could also consider the $3-4-5$ model. The vacuum polarization for
this model becomes
\be
\Pi_{\mu \nu}^{(3-4-5)} (p) = \frac{e^2}{\pi}
\frac{(3^2+4^2+5^2)}{2} \left( \delta_{\mu \nu} - \frac{p_\mu
p_\nu}{p^2} \right)
\ee
and hence a mass gap $\mu^{(3-4-5)} = 5 \frac{e}{\sqrt{\pi}}$.
It is easy to see that the fermion number current of the model,
defined by
\begin{equation}
J_\mu^F = J_\mu^{R,3}+J_\mu^{R,4}+J_\mu^{L,5}
\end{equation}
is anomalous:
\begin{equation}
\partial^\mu J_\mu^F (x) = i(3+4-5) \frac{e}{4 \pi}  \epsilon^{\mu \nu}
\partial_\mu A_\nu (x)  .
\end{equation}
This result agrees with a previous calculation in reference\cite{Ao:Hi:93}.

\section{Discussion}
We have shown that the KNN scheme for implementing chiral fermions
passes a simple perturbative test in 2+1 dimensions.
Our perturbative scheme renders the effective action finite after
the subtraction of effective actions with homogeneous mass terms.
The gauge variant term in the effective action corresponds
exactly to the consistent anomaly: in contrast to other regularization schemes,
the real part of the effective action {\em is} gauge invariant.
To obtain this result, we made  the infinite $s$ summation
well-defined by restricting the range of the gauge interaction.
This restriction of course breaks gauge invariance, in both  the
real and the imaginary part of the effective action, but when the
range of the  gauge interaction is taken to infinity gauge
invariance is restored in the real part and broken only in the
imaginary part, giving the correct anomaly.

A rather more difficult test of the KNN scheme is a perturbative calculation in
$4+1$ dimensions. We expect the scheme to work just as well
in $4+1$ dimensions as in $2+1$, but of course an explicit demonstration is
necessary. If this test is passed, we would expect the KNN
regularization of more complicated anomaly-free chiral gauge theories like the
Standard Model to also have the correct continuum limit.

The infinite extra dimension that
is needed to make a truly chiral fermion has been shown here to
be tamable in perturbation theory: Narayanan and Neuberger have
also given a finite and hence computable non-perturbative
effective action, in the form of an overlap. One obvious research
goal is to
produce a version of the non-perturbative overlap formula
\cite{Na:Ne:94} that can be used in practical Monte-Carlo type calculations.

S.A. is supported in part by the Grants-in-Aid of the Japan Ministry of
Education(No. 06740199).
R.B.L is indebted to R. Narayanan and H. Neuberger for
providing invaluable input and encouragement.

\vspace{1cm}
\Large \noindent {\bf Appendix}
\normalsize
\vspace{1cm}

The
effective action for an odd number of gauge fields vanishes, by Furry's
theorem. This holds not only in the continuum\cite{I&Z} but also on the
lattice, where we have to worry about graphs with seagull
vertices. The point is that the charge conjugation matrix $C$,
defined by $C \gamma_\mu C^{-1} = - \gamma_\mu^T$, transforms the
propagator like $C S(q) C^{-1} = S^T (-p)$ and the $m$-gauge
field vertex factor
 like $C \partial_\mu^m S^{-1}(p) C^{-1} = (-1)^m \partial_\mu^m
(S^{-1}(-p))^T$, where $T$ denotes the transpose in spinor space.
So regardless of the number of seagull vertices, it is easy to
show (by insertion of factors $CC^{-1}$) that a given graph
with a fermion loop and $n$ attached gauge fields is equal to the
same graph with the reverse orientation, up to a sign $(-1)^n$.
Thus for odd $n$ the two orientations cancel. The only
complication we have blithely skipped over is the action of the charge
conjugation matrix on the chiral ``mass'' term $M(p)P^R +
M^\dagger (p) P^L$ in the propagator. For a
lower dimensional ``target'' space of dimension $d-1=2,6,10, \cdots$,
we find that $C \gamma_5 C^{-1} = - \gamma_5^T$. We still get
$C S(q) C^{-1} = S^T (-p)$, but now $T$
denotes transposition in $s$--space {\em as well as} Dirac space.  Of
course this is exactly what we need to transform the graph (say
the one in
figure 3, with $n$ odd) into the graph with reverse
orientation, up to a sign $(-1)^n$.
For target dimension $d-1=4,8,12,\cdots$,
we find that $C \gamma_5 C^{-1} = \gamma_5^T$, so the transformed
graph is equal to the graph with reverse orientation, up to a
sign $(-1)^n$ {\em and} a transposition of each propagator in
$s$--space. But since we trace over $s$, the graph is invariant
under an $s$--transposition of each propagator and we are done.

\vspace{1cm}
\Large \noindent {\bf Figure Captions}
\normalsize

\begin{enumerate}
\item An $n$-photon vertex.
\item Graphs contributing to the vacuum polarization. (a) The non-seagull
graph.
(b) The seagull graph.
\item A fermion loop with $n$ photons attached.
\end{enumerate}


\begin{thebibliography}{99}
\bibitem{Fr:Sl:93}
S. A. Frolov and A. A. Slavnov, Phys. Lett. { B309} (1993) 344 .
\bibitem{Fa:Sl:89}
L. D. Faddeev and A. A. Slavnov, {\em Gauge Fields. Introduction to Quantum
Theory}, Second Edition (Benjamin, Reading, Massachsetts, 1989).
\bibitem{Ka:Sm:81}
L. H. Karsten and J. Smit, Nucl. Phys. { B183} (1981) 103 .
\bibitem{Ni:Ni:81}
H. B. Nielsen and M. Ninomiya, Nucl. Phys. { B185} (1981) 20 .
\bibitem{Na:Ne:93}
R. Narayanan and H. Neuberger, Phys. Lett. { B302} (1993) 62 .
\bibitem{Ka:92}
D. B. Kaplan, Phys. Lett. { B288} (1992) 342 .
\bibitem{Go:et:al:93-94}
M. F. L. Golterman, K. Jansen, D. N. Petcher and J. C. Vink,
Phys. Rev. D49 (1994) 2604; Nucl. Phys. B (Proc. Suppl.) 34
(1994) 593; M. F. L. Golterman and Y. Shamir, Wash. U. preprint
HEP/94-61 (1994) (hep-lat/9409013).
\bibitem{Na:Ne:94}
R. Narayanan and H. Neuberger, Nucl. Phys. { B412} (1994) 574 .
\bibitem{Pe:88}
A. Pelissetto, Ann. Phys. 182 (1988) 177.
\bibitem{Bo:Ko:87}
G. T. Bodwin and E. T. Kovacs, Phys. Rev { D35} (1987) 3198 .
\bibitem{Na:Ne:94b}
R. Narayanan and H. Neuberger, Nucl. Phys. { B} (Proc. Suppl.)
{ 34} (1994) 587 .
\bibitem{AK}
S. Aoki and Y. Kikukawa, Modern Physics Letters { A8} (1993) 3517 .
\bibitem{Fu:94}
K. Fujikawa, University of Tokyo preprint TU-678 (1994).
\bibitem{Co:Lu:89}
A. Coste and M. L\"{u}scher, Nucl. Phys. { B323} (1989) 631 .
\bibitem{So}
H. So, Prog. Theor. Phys. { 73} (1985) 585; { 74} (1985) 528 .
\bibitem{Na:Ne:93b}
R. Narayanan and H. Neuberger, Phys. Rev. Lett. { 71} (1993) 3251 .
\bibitem{Ao:Hi:93}
S. Aoki and H. Hirose, Phys. Rev. { D49} (1994) 2604 .
\bibitem{I&Z}
C. Itzykson and J-B Zuber, {\it Quantum field theory}, (McGraw-Hill, New York,
1980), p 276.
\bibitem{Ja:Ra:85}
R. Jackiw and R. Rajaraman, Phys. Rev. Lett. { 54} (1985) 1219 .
\bibitem{Ha:Ts:87}
K. Harada and I. Tsutusi, Phys. Lett. { B183} (1987) 311 .
\bibitem{Ao:88}
S. Aoki, Phys. Rev. Lett. { 60} (1988) 2109 ; Phys. Rev. { D38} (1988) 618 .
\bibitem{Fu:Ka:88}
K. Funakubo and T. Kashiwa, Phys. Rev. Lett. { 60} (1988) 2113 .
\bibitem{Ki:et:al:88}
T.D. Kieu, D. Sen and S.-S. Xue, Phys. Rev. Lett. { 61} (1988) 282 .
\end{thebibliography}
\end{document}